# A successive sub-grouping method for multiple sequence alignments analysis

Stefano Marino
Department of Biology, University of Padova
Via Ugo Bassi 58b, 35121, Padova
(Dated: May 30, 2007)

A novel approach to protein multiple sequence alignment is discussed: substantially this method counterparts with substitution matrix based methods (like Blosum or PAM based methods), and implies a more deterministic approach to chemical/physical sub-grouping of amino acids. Amino acids (aa) are divided into sub-groups with successive derivations, that result in a clustering based on the considered property. The properties can be user defined or chosen between default schemes, like those used in the analysis described here. Starting from an initial set of the 20 naturally occurring amino acids, they are successively divided on the basis of their polarity/hydrophobic index, with increasing resolution up to four level of subdivision. Other schemes of subdivision are possible: in this thesis work it was employed also a scheme based on physical/structural properties (solvent exposure, lateral chain mobility and secondary structure tendency), that have been compared to the chemical scheme with testing purposes.

In the method described in this chapter, the total score for each position in the alignment accounts for different degree of similarity between amino acids.

The scoring value result form the contribution of each level of selectivity for every individual property considered. Simply the method (called M_Al) analyse the n sequence alignment position per position and assigns a score which have contributes by aa identity plus a composed valuation of the chemical or of the structural affinity between the n aligned amino acids.

This method has been implemented in a series of programs written in python language; these programs have been tested in some biological cases, with benchmark purposes.

INTRODUCTION

An clear cut subdivision of amino acids (aa) into subsets defined on the basis of chemical/physical characteristics is not a feasible task. Several chemical and physical properties differentiate these molecules from each other, such as, polarity, dimension, functional group, mobility of lateral chain and so on. In the past, the attempts to divide aa into groups that takes into account some of these different properties (8, 17, 18, 19) resulted in a scheme characterized by the superimposition of different subsets, like the Venn diagram reported in figure 1. In the latter scheme many intersection can be observed among the different subsets. These areas of overlaps hinder a rigorous treatment of the problem.

Currently, a simplification of the derivation scheme is normally used, leading to some of the diffused scheme of chemical/physical properties derivation, like the simplified Venn diagram used by ClustalW, illustrated in figure 1.

Another way, very useful and effective, to treat aa conservation in a protein multi-alignment implies the use of matrixes built by taking into account the frequencies of aa substitution.

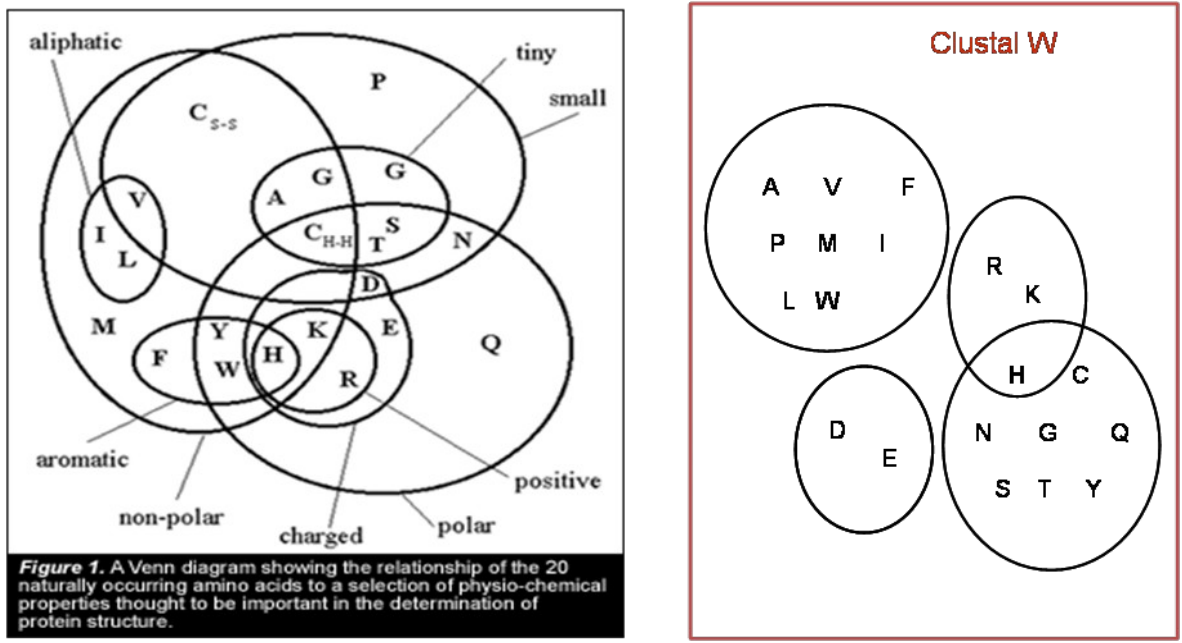

**Figure 1:** left: A Venn diagram showing the relationship of the 20 naturally occurring amino acids to a selection of physical/chemical properties thought to be important in aa diversification.
Right: Diagram by which ClustalW sub-groups the 20 aa into subset defined on the basis of affinity of their chemical physical properties (this scheme bring to the assignment of the symbols " **\*** ", " **.** ", " **:** " ; see also references 1and 12)

These knowledge-based methods allow to define substitution matrixes, where for each position ($a_{i,j}$) a value defined a sort of "probability" (more exactly accounts for a frequency) for the $aa_i$ to be substituted with $aa_j$. These values derive from a dataset of aligned sequences. The procedures by which the sequence alignment is composed and treated (for example, clustered or not) and by which pair wise scores are assigned to $aa_i >$ $aa_j$ (substutution of the $aa_i$ into $aa_j$ ) are specific for different methods. The most famous, and widely employed, methods of substitution matrix are PAM and BLOSUM (4,5).
In the alternative procedure described in this work, a method based on user defined scheme of derivation is presented ; the method has been implemented in a set of computer programs written in Python language (16). The algorithm, called M_Al, is substantially based on the derivation of a scheme of amino acidic sub-grouping on the basis of chemical or physical properties of amino acids.

**METHODS**

While designing the method, different "schemes of derivation" were drawn. Initially, they were all tested in comparison to ClustalW and JalView (in particular its alignment "quality" scores, see references 2, 14), and the results of this preliminary benchmark (not shown) allowed a selection of the two principal schemes of derivation shown in figure 2 and figure 3, the former aiming to derive more chemically characterized information from the alignment the latter to extract more structurally derived information.
The use of the first scheme of derivation (Fig 2) in the method was called Ch:M_Al (Chemical Multi Alignment), while the use of the other, more structurally/physically oriented , was called Ph:M_Al (Physical Multi Alignment, scheme shown in figure 3).
The rational behind the "chemical" sub-grouping scheme in figure 3 is: derivation from an initial unique set of aa (set **I**) for lateral chain (R) polarity/functional group OR for lateral chain hydrophobicity/aromaticity.

With three successive "derivation", according to the scheme, it was possible to obtained the sub-grouping depicted in figure 3. The sub-groups obtained are three, labelled as **II** (**a** and **b**), **III** (from **a** to **d**) and **IV** (from **a** to **h**)

As to the scheme of derivation depicted in figure 3, a 'physical sub-grouping' (yet more structurally informative) was derived with three different scheme of sub-grouping (three separated successions for three different partial derivation, each one for a different property, i.e. R mobility, solvent exposure, secondary structure propensity).

The structure of the method and of its formalization make it is possible to define and describe a "expected" associated scoring function, which derived from the occurrence of the i-th letter in each subset (see later).

The useful aspect of these two types of different grouping schemes is that they aim to account separately for the chemical aspects of the amino acidic conservation and the structural ones.

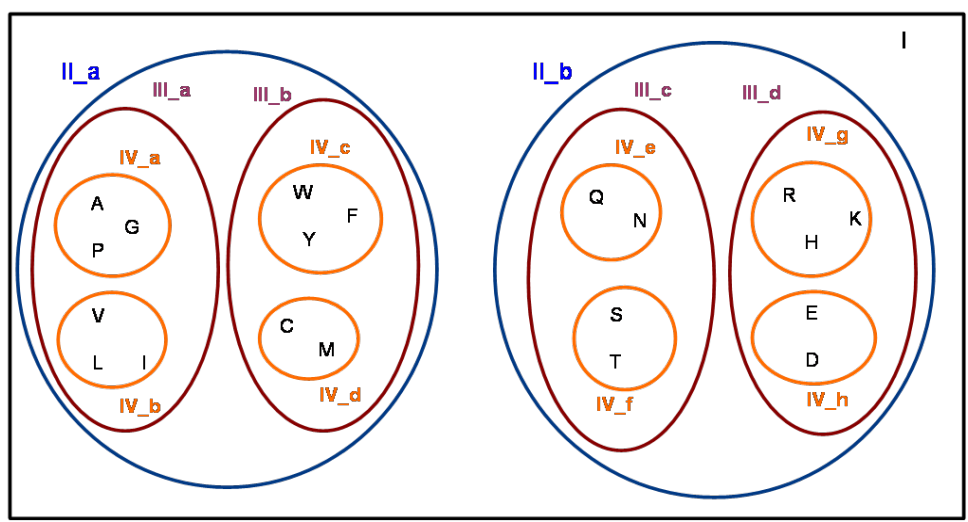

**Figure 2**: The scheme used for chemical sub-grouping (i.e. chemical scheme of derivation) has been drawn taking into account polarity/apolarity or Functional group/aromaticity. From an initial set, pointed as **I**, (contoured with black line), after the first derivation two subsets, pointed as **II**, are formed; then after a successive derivation of each of the two II subset for the same "property" four subsets, pointed as **III**, are drawn; finally from each of the four III-subsets derive the eight subsets pointed as **IV**.

It has to be mentioned that the sub-grouping method and its implementation can work also with other scheme of derivations, different and alternatives to the ones purposed in this work; in particular the idea of design schemes by rigorously deriving for a single and specific property (for example, the derivation for lateral chain mobility depicted in figure 4) can be also more accurate from an analytical point of view.

A schematic description of the algorithm used follows.

Given n aligned sequences as input:

1) one of the derivation schemes presented above (or implementation of a user defined one) is chosen

2) transformation of each aligned word into another word depending on the vocabulary of each subgroup (i.e. each aligned aa sequence is transformed into m sequences for each m order of derivation):

For instance, considering the first derivation of 20 aa (group indicated in figure 2 as I) into 2 subgroup of 9 Polar aa and 11 Apolar aa (indicated in figure 2 respectively as II_b and II_a), letters of vocabulary for this derivation order (i.e. first) is II_a and II_b, and according to the defined scheme of derivation in figure 2, each aligned word is translated into II_a or II_b.

The n aligned sequence are all transformed in this way and then analyzed.

3) For each alignment obtained (one for each derivation characterized by its own vocabulary) the conservation is calculated (i.e. the word conservation for each column of the alignment, for each of the m transformations)

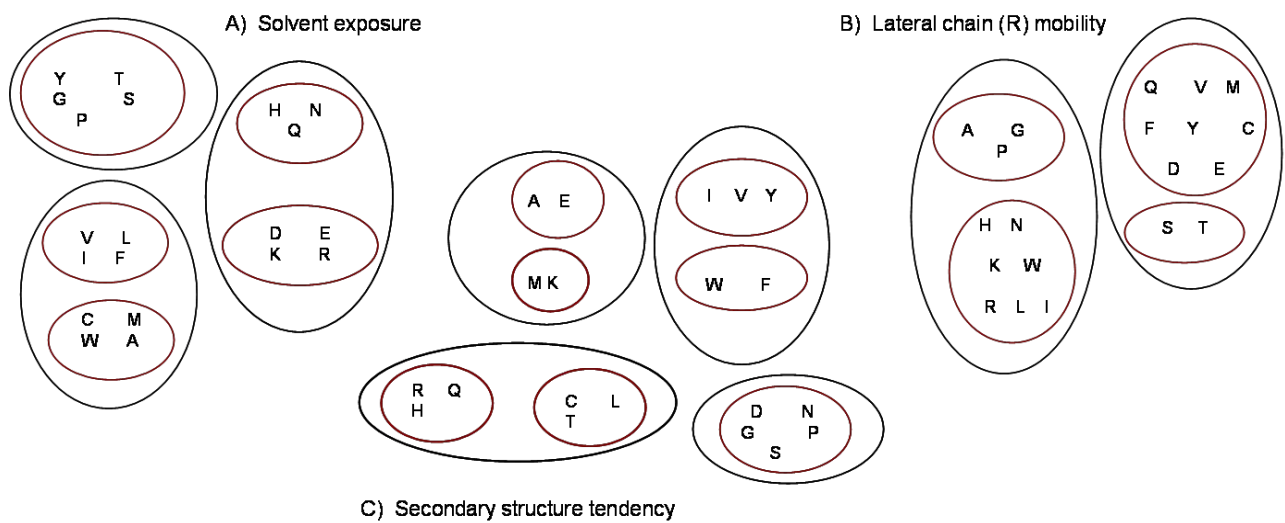

**Figure 3**: Physical/structural subdivision of aa. The properties analyzed, i.e. solvent exposure (7), secondary structure tendency (6, 15) and lateral chain mobility (12), are considered independent property. For this reason the three schemes depicted in figure are used contemporary. The method M_Al with this derivation scheme implemented is called Ph:M_Al. The form of the scoring function (*Sc*) is that of Ch:M_Al (see text description), with the three contributes that are linearly summed: *Sc*(Ph:M_Al) = *Sc* (solvent exposure) + *Sc* (R mobility) + *Sc* (Secondary structure tendency). As the form of the scoring function is that of Ch:M_Al, the same happens for the expected associated polynomial (*Se*, see text). *Se* in this case is simply the sum of three polynomial from each of the three independent schemes.

4) Then for each aligned position, a scoring function (Sc) is evaluated in the form :
$$Sc = I_0(aa) + f(aa) + f_x^1(aa) + f_x^2(aa) + \ldots + f_x^m(aa)$$
Where $I_0(aa)$ is the starting value of aa identity, m is the number of derivation (m = 3 in the case of chemical conservation scheme depicted in figure 3, that brings aa grouping from set I to subset IV), and x is the property by which the partial derivation is made. However in our cases x is not a rigorous property for both schemes, but a practical determinant of subdivision: x for Ch:M_Al is a sort of polarity/functional group OR apolarity/aromaticity index, x for Ph:M_Al are three different properties (solvent exposure, secondary structure tendency and lateral chain mobility ) linearly summed up (see also text in figure 2 and figure 3).

Turning back to the general explanation of the method, a consideration is due, f(aa), as previously defined, scores always 100% of conservation, so its contribute to the function is set to be 0 and consequentially it is not informative (and could be droped from the evaluation of the scoring function) . So the scoring function become:
$$Sc = I_0(aa) + f_x^1(aa) + f_x^2(aa) + \ldots + f_x^m(aa)$$
$f_x^j$ (with $1 < j \leq m$) is the function of conservation calculated after each element of the column has been transformed according to the j-order derivative is applied ($f_x^1$ is first derivative and so on);
In other words, referring to figure 2, the scoring function sum up all the scores obtained for each level of sub-grouping (I, II, III, IV drawn in the figure) plus the term corresponding to amino acidic conservation ($I_0(aa)$).
5) Let the amino acidic identity ($I_0(aa)$ ) for that column of the alignment be equal to $I_0(aa) = k/n$, where k

is the number of occurrence of the element Y in the column.

Then an expected polynomial function (Expected score, Se ) of the same order of the Sc scoring function is derived, based simply on expected frequencies of occurrence by which the (n-k) not-Y positions can be similar by chance. This operation is made for all the transformed sequences, so evaluating the expected values for each derivative step (i.e. for each transformed column is evaluated the number of occurrence by chance of a certain word in the (n-k) positions of the column).

The Se scoring function will be:

$$Se = I_0(aa) + E_x^1(aa) + E_x^2(aa) + E_x^3(aa) + \ldots + E_x^m(aa)$$

Where $E_x^j$ (with $1 < j \leq m$) is the function of expected conservation calculated after each element of the column has been transformed according to the j-order derivative is applied

6) For each position: comparison between the calculated value Sc and Se, and assignment of a "+" (up conservation of the property x) or " - " if the calculated value is, respectively, higher or lower than expected value; a total calculated score and a total expected score are written and saved in the output file

7) In the output files are also saved other computed values, among which a normalized (from 1 to 10) list of value that give a simple scoring value for each position (figure 5)

The advantages of the method are a good resolution of the alignment conservation (for the property x by which the scheme of derivation is defined) given by the form of the scoring function; and then the possibility to compare in the same run of calculations the two polynomial functions (Sc and Se) highlighting in the output file the significant up-conserved ("+") or down-conserved ("-") positions of the alignment (figure 5). Also unique values, integrating the values for each position for all the aligned aa, are given by the programs: this values can be useful for an overall scoring of an alignment (see later).

A simple example can better illustrate the principle of the method; referring to physical scheme of derivation for lateral chain mobility (drawn in figure 4), chosen for sake of clarity.

Given a list of 8 aligned amino acid A P P L G H P E and applying the chosen scheme of derivation the alignment turns into: 1) I I I I I I I I , 2) II-a II-a II-a II-a II-a II-a II-a II-b, 3) III-a III-a III-a III-b III-a III-b III-a III-c.

These transformed columns are analyzed for the conservation, and each calculated value is inserted in the polynomial scoring function presented above, giving the following score:

$Sc = 1.875$

Normalizing for the maximum possible value of the scoring function (i.e the value corresponding to a full aa conservation): $Sc = 0.625$

As to the associated expected function (Se): considering that 3/8 represents the maximum identity value, $I_0(aa)$, for that aligned position, it is possible to calculate the expected associated score (Se)

$Se = 0{,}375 + 0{,}056 + 0{,}187 = 0{,}618$

From comparison of the two values, a trend is derived of "over conservation" of the property (i.e. a trend to maintain a kind of R mobility, either higher or lower) for that position (labelled as + in the output string); note that if the calculated value is higher than expected, it can be argued that in the position a constraining force acts on the property rather than on a specific amino acid selection: so the best range of resolution of the method is when aa conservation is not so high, but calculated value is clearly higher than expected one. In the example proposed a certain trend of over conservation of the properties over the pure amino acid conservation is detectable from the method, even if the method clearly work much better if number of aligned sequences is larger. Using the substitution matrix based methods available in the literature it is possible to obtain information only on the level of conservation (in some cases with a scoring function very resolved), and not a tendency (and a quantification of this tendency) of the property conservation, given the initial aa identity.

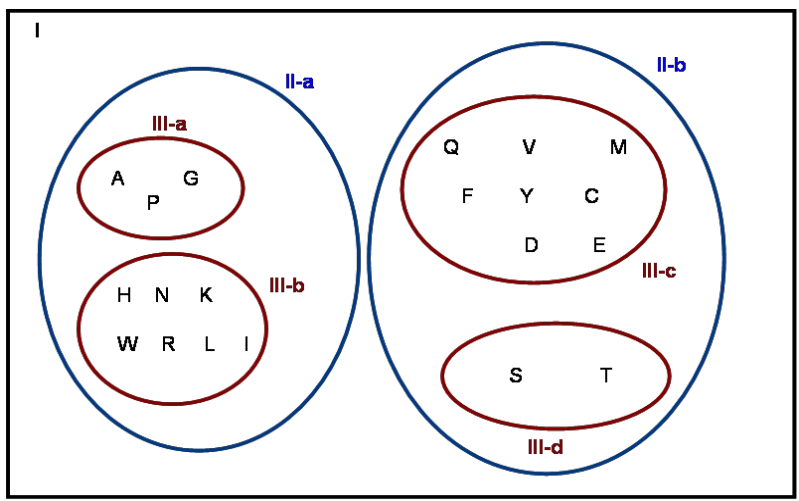

**Figure 4:** A detailed representation of the scheme of derivation of lateral chain mobility (using the rotamer library of Deep View 3.7, see references 11 and 12) yet presented in figure 4. This scheme alone is presented to simply illustrate, with graphical support, the example on how the method works presented in the text.

## RESULTS AND APPLICATIONS

**Comparisons with ClustalW in the case of Hemocyanins**
As a first brief example of application, the study of chemical and physical properties distribution evaluated with the both Ch:M_Al and Ph:M_Al in the case of the type-III copper proteins Hemocyanins (Hc) is presented.
Hcs have a complex quaternary structure (20,22, 23), but substantially are made up of a number of different monomers (Functional unit, FU) that are in number of 7 in the case of *O. dofleini*; these FU shared a "low" conservation in all Hcs, with an average of 43% of aa identity for *O. dofleini* proteins.
As all other Hcs the single FU is a peptide (in the example of figure 7 it is visualized to the 1odg FU, with crystallographic code 1js8, 23) that can be divided into two well detectable domains: one with a main alpha helix structural characterization (N-terminal domain, of about 30 KDa) and the other with mainly a Beta strand characterization (C-term domain, of about 20 KDa). The two domains are in contact with a well defined interface surface, characterized by extensive hydrophobic and electrostatic interactions (20,21).
Given the ClustalW-made multiple alignment of the 7 functional units (FUs) of *Octopus dofleini* respiratory protein Hemocyanin (Hc, figure 5), the alignment file was submitted to M_Al.
An overall good accordance is shown between ClustalW (CW) conservation assignments and M_Al (both Ch:M_Al and Ph:M_Al) scores: all conserved positions detected by CW were also detected by M_Al, as it can be seen in the example shown in figure 5.
Particular attention was paid to the interface region between the 30 KDa and the 20 KDa domains.
The outstanding observation is that while looking for contact regions of the two domains (blue coloured region in the right part of figure 6) a clear conservation of chemical properties is detectable (Figure 6, middle part of the figure).
As to this interface region, the result for the structural analysis (Ph:M_Al) is different, not indicating a high conservation of the structural properties in the interface regions (figure 6, left part of the picture).
Moreover a tendency of chemical "up-conservation" (i.e. chemical conservation calculated values higher than expected) relative to the contact region can be quantified with a value of about + 6% rather than expected, meaning that not also the region is chemically conserved, but also higher than expected. This seems to indicate a considerable selection acting on chemical properties of aa at the interface, portion of the protein that is thought to be strictly involved in Hc functionality, among all preventing Hc from undesirable type-3 copper specific reactivity (20,21).

Difficulties in attempts to separate the 2 domains without denaturating the whole protein are described in literature (20, 21); as previously mentioned these are thought to be due to extended hydrophobic and electrostatic interactions: accordingly Ch:M_Al detects higher conservation values, while neither Ph:M_Al or ClustalW or JalView (data not shown) computed high scores for these interface positions.

**Comparison with ClustalW/jalView in the case of carbonic anhydrases (CAs)**

In this part of the work an investigation with benchmark purposes was made on two different phylogenetic grouped carbonic anhydrase with cytosolic activity

It was chosen to compare M_Al results with JalView (JV) "quality score" method (2, 14).

This method output a "quality alignment score" which is calculated for each column in an alignment by summing, for all mutations, the ratio of the two BLOSUM 62 scores for a mutation pair and each residue's conserved BLOSUM62 score (4, 13,14).

The main goal of this part of the work are: first to compare M_Al calculations with those of a well known substitution matrix based method, as a sort of test to asses the differences between the outputted values; a second goal was to test the utility of the expected values associated to the calculated ones.

In particular while in figure 5 are shown the position of up or down conservation (meaning that calculated conservation is higher or lower than expected) of the chemical or structural properties, in this part of the analysis the overall numerical values of the scoring function (Sc) and of the expected function (Se) are compared.

The taxon specific alignment refers in particular to mammalian cytosolic CAs (CAI, CAII, CAIII, CAVII) and vertebrate CAs (all complete sequences retrievable with the "protein" search at NCBI site, http://www.ncbi.nlm.nih.gov/).

When mammalian cytosolic CAs (average value for amino acidic identity of 60 %) are submitted to the programs, the obtained results are:

JV score = 0.743

Ch:M_Al calculated score (Sc) = 0.815

Ch:M_Al expected score (Se) = 0.775

Ph:M:Al Sc = 0.810

Ph:M_Al Se = 0.792

A trend of about 5 % over the statistical value, indicates a conservation of chemical properties higher than expected: in other words a trend of selection on chemical properties is revealed.

Also structural conservation is comparably good, even if the difference between expected and calculated values is lower.

The same calculations were made for vertebrate high activity cytosolic CAs (with an average value for amino acid identity of 58 %) giving the following results:

JV = 0.741

Ch:M_Al calculated score (Sc) = 0.791

Ch:M_Al expected score (Se) = 0.761

Ph:M_Al Sc = 0.791

Ph:M_Al Se = 0.779

In CAs case the outstanding result is that JV (using Blosum62 matrix) do not operate a clear discrimination for cytosolic mammalian CAs and vertebrate CAs; in contrast both Ch:M_Al and Ph:M_Al detect a higher conservation for mammalian CAs; however a little more significative is the overall chemical conservation calculated for mammalian proteins (0.81 vs 0.79).

If we look to the differences between expected and calculated values we see a light trend of up conservation (calculated values higher than expected) both for Ch:M_Al and Ph:M_Al,; again this behaviour is more pronounced for chemical conservation in mammalian CAs.

A visualization of these results is presented in Figure 7.

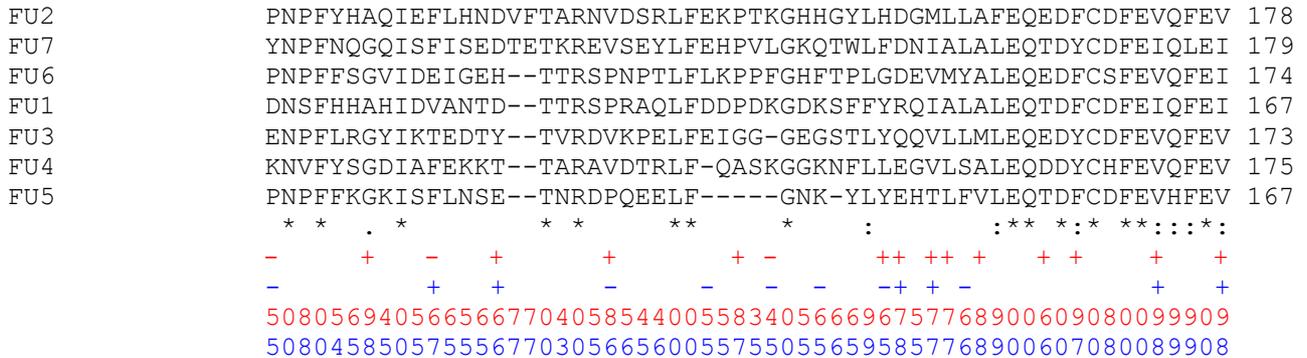

```
FU2        PNPFYHAQIEFLHNDVFTARNVDSRLFEKPTKGHHGYLHDGMLLAFEQEDFCDFEVQFEV 178
FU7        YNPFNQGQISFISEDTETKREVSEYLFEHPVLGKQTWLFDNIALALEQTDYCDFEIQLEI 179
FU6        PNPFFSGVIDEIGEH--TTRSPNPTLFLKPPFGHFTPLGDEVMYALEQEDFCSFEVQFEI 174
FU1        DNSFHHAHIDVANTD--TTRSPRAQLFDDPDKGDKSFFYRQIALALEQTDFCDFEIQFEI 167
FU3        ENPFLRGYIKTEDTY--TVRDVKPELFEIGG-GEGSTLYQQVLLMLEQEDYCDFEVQFEV 173
FU4        KNVFYSGDIAFEKKT--TARAVDTRLF-QASKGGKNFLLEGVLSALEQDDYCHFEVQFEV 175
FU5        PNPFFKGKISFLNSE--TNRDPQEELF-----GNK-YLYEHTLFVLEQTDFCDFEVHFEV 167
            * * . *         * *    **     *     :    :** *:* **:::*:
           -    +  -  +       +         + -         ++ ++ +   + +     +   +
           -        +  +      -         - -   -    -+ + -             +   +
           5080569405665667704058544005583405666967577689006090800999 09
           5080458505755567703056656005575505565958577689006070800899 08
```

**Figure 5:** An example of score for M_Al method compared with the evaluation of ClustalW chemical affinity for each column of the alignment (with code *=total conservation, : and . for certain chemical/physical conservation, blank space for non detectable conservation of chemical/physical properties).

In this example only a portion of the region connecting the 2 domains for *Octopus dofleini* FUs alignment is presented; just for giving an idea of a typical output of the program M:Al; the potential "chemically" OR "structurally" up- ("+") and down- ("-") conserved positions (meaning that calculated values for each column of the alignment are higher, "+", or lower,"-", than expected values, see text description) are signed at the end of the computations.

In red chemical score (normalized from 1 to 10, with "0" standing for "10") of the M_Al method (Ch:M_Al) are presented, in blue the structural scores (Ph:M_Al).

From a detailed analysis of Figure 7 it is possible to note that some differences are present while looking position per position, but the overall affect is to have a strong selection for maintaining the general chemical and structural features of the protein.

Finally the considerable agreement between the Ch:M_al, Ph:M_al and Jalview has to be noted.

Taking into account the difference between the derivation of the methods, their relevant agreement detectable in CAs analysis can be hypothetically explained with a high number of positions that are evolutionary strongly constrained, resulting in a very small range of possible mutations.

This little range of mutation forces amino acid substitutions to be between very similar aa, in this way substitution are often classified in the same way (i.e. "consensus" mutations) by the three so differently derived methods.

However some differences persist (figure 7), and give indications of the utility of the methods, especially highlighting the utility and gain in critical information obtainable with the contemporary usage of different derived methods.

## CONCLUSIONS

Summing up the results of this analysis, the method M_Al shows good overall agreement with ClustalW/JalView (figure 5, 6, 7) but at the same time allows user-specific defined scheme of derivation which provide the ground for a more critical analysis of the results obtained.

A "correct" scheme of derivation can give a quite detailed picture of the distribution of the considered property/ies (like polarity/hydrophobicity) in the multi-alignment as well as an estimation of its significance.

Two aspects of the method can be of some utility:

1) choosing an opportunely defined chemical scheme (like that used in Ch:M_Al, see figure 2) of derivation and a more structurally/physical oriented scheme (like that used in Ph:M_Al, see figure 3), it is possible to separate two mayor contributes to chemical/physical conservation, thus eventually discerning which of the two results more determinant; this possibility is well shown by the Hemocyanin example

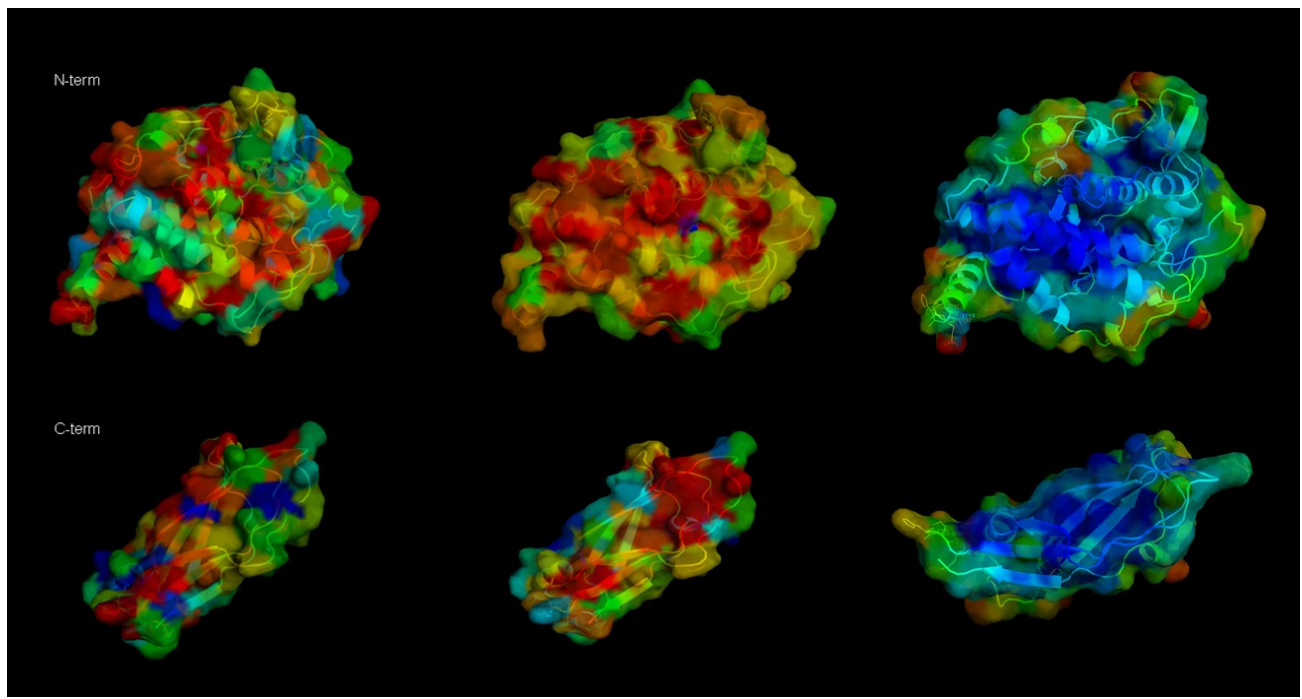

**Figure 6**: left, Ph:M_Al scores plotted on molecular surface of hemocyanin N-terminal sub-unit (up) and C-terminal subunit (down); middle, Ch:M_Al scores plotted on the same surface type; right, cystallographic structure for *O.dofleini* Hc (pdb code 1js8, 23) coloured by B-factor distribution. Orientation of the structures are chosen for enhance the conservation in the interface regions, by which the two sub-units (30 KDa for the N-terminal and 20 KDa for the C-terminal) result to be in contact, as it can be seen looking at B-factor distribution (colurs range from blue, for low temperature factor, to red, for high temperature factor). In figure, referring to B factor distribution, the interface region between 30 KDa and 20 KDa subunits is coloured in blue because of the stabilizing effects of the interactions (20,21), which have a correspondence on crystallographic B-factors (22). For images referring to Ph:M_Al and Ch:M_Al scores, the colouring code ranges from red (higly scoring, or higly conserved positions) to blue (poorly scoring positions), passing through orange, yellow, green and cyan. From the picture it is evident a clear better conservation of "chemical" properties in the interface region, in respect to physical/structural properties conservation.

2) making use of the "calculated" and "expected" polynomial function (respectively Sc and Se, previously described) it is possible to "weight" the conservation score in the multi alignment, gaining more information on the significance (with the plus of the chemical and structural separated information) of the score.

The latter has several implications: one regards the significance of the overall score on the multi alignment and this is well explained in the CAs analysis.

In general some applications of M_Al method can be linked to comparisons with validation purposes with substitution matrix based alignment and related purposed phylogenetic distances; due to the different nature of the Blosum or PAM matrix and the M_Al schemes a good validation can be obtained in the cases in which the output are consistent one each other (as in the Cas case presented here).

Another implication is the possibility to use the method with protein engineering purposes.

Positions with high chemical conservation (good scoring value and higher than expected) can be considered as potential positions to be mutate with a similar aa (in accordance to the used scheme) , while position with a significantly lower score that expected, probably need more attention before a "consensus" mutation is planned (given the high dependency in the alignment to the specific aa rather than to the specific property); many others structural information on hotspots of a protein structure can be extracted from the multi-alignment with other strictly related proteins.

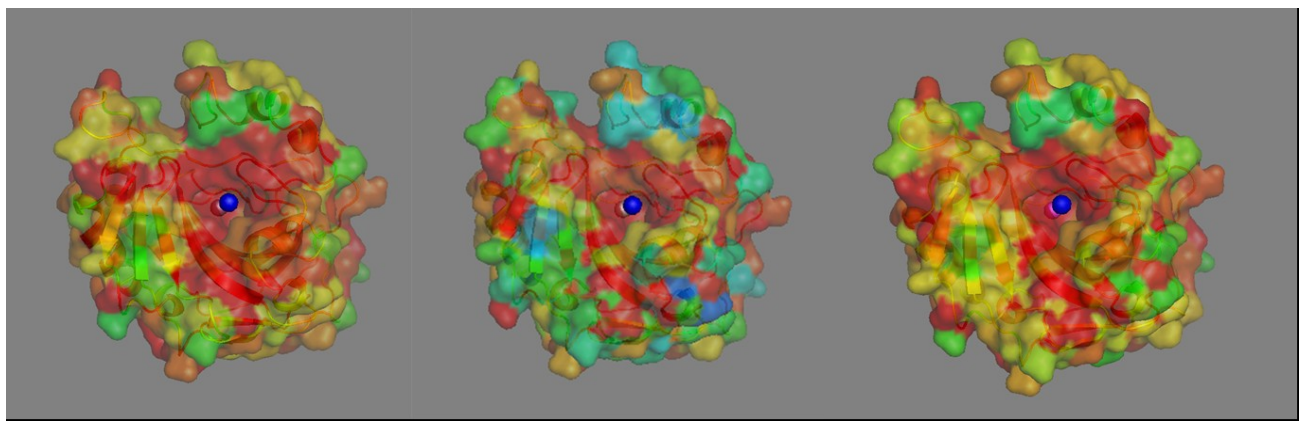

**Figure 7:** Ch:M_Al , JV, Ph:M_Al (from left to right) results visualized with PyMol (http://pymol.sourceforge.net/) for vertebrates cytosolic CAs. JV results refer to "quality" score ( 2, 14).
This PyMol visualization is obtained with a script in Python that inserts the calculated value obtained from M_Al and JV in the pdb column relative to B-factors. Thus code colour is that of B-factor for PyMOl, with higher value (that in our case mean higher conservation) coloured in red and lower value coloured in blue. Intermediate value last respectively from orange to cyan (i.e. from good to poor conservations). It's the same code colour described in figure 6. Orientation of proteins is with enzymatic cleft pointing out from the page.